\documentclass[conference]{IEEEtran}
\IEEEoverridecommandlockouts
\newcounter{paperpages}
\AddToHook{shipout/after}{\stepcounter{paperpages}}
\AddToHook{enddocument/afterlastpage}{%
  \ifnum\value{paperpages}>6\relax
    \PackageWarningNoLine{camera-ready}{Six-page limit exceeded:
      \arabic{paperpages} pages. Shorten the manuscript before submission}%
  \fi
}
\usepackage[utf8]{inputenc}
\usepackage[T1]{fontenc}
\usepackage{cmap}
\usepackage{cite}
\usepackage{amsmath,amssymb,amsfonts}
\usepackage{graphicx}
\usepackage{textcomp}
\usepackage[table]{xcolor}
\usepackage{booktabs}
\usepackage{subfigure}
\usepackage{algorithm}
\usepackage{algpseudocode}
\usepackage{amsmath,amssymb} % para símbolos matemáticos
\usepackage[protrusion=false]{microtype} %solve gutter problem
\usepackage{tikz}
\usepackage{graphicx}
\usepackage[nolist]{acronym}
\usepackage{fontawesome}
\usepackage{makecell}
\usepackage{pgfplots}
\pgfplotsset{compat=1.17} 
\usepgfplotslibrary{groupplots}
\usetikzlibrary{fit,positioning,calc,arrows.meta}
\usetikzlibrary{arrows.meta, positioning, calc, shapes.geometric, fit, backgrounds}

\def\BibTeX{{\rm B\kern-.05em{\sc i\kern-.025em b}\kern-.08em
    T\kern-.1667em\lower.7ex\hbox{E}\kern-.125emX}}
\begin{document}

%\title{An Agentic-based Control-Loop for Stable QoS Enforcement in Mobile Networks}
\title{Agentic Orchestration for SLA-Constrained Network Slicing in Programmable Mobile Data Planes}

\author{\IEEEauthorblockN{Rodrigo Moreira\textsuperscript{1}, Larissa Ferreira {Rodrigues Moreira}\textsuperscript{1}, Joberto S. B. Martins\textsuperscript{2},\\ Tereza C. Carvalho\textsuperscript{3},
Flávio de Oliveira Silva\textsuperscript{4}}
\IEEEauthorblockA{
\textsuperscript{1}Federal University of Viçosa (UFV), Minas Gerais, Brazil\\
\textsuperscript{2}University of Salvador (UNIFACS), Salvador, Brazil\\
\textsuperscript{3}University of São Paulo (USP), São Paulo, Brazil\\
\textsuperscript{4}University of Minho (UMinho), Braga, Portugal\\
Emails: rodrigo@ufv.br, larissa.f.rodrigues@ufv.br, joberto@ieee.org, terezacarvalho@usp.br, flavio@di.uminho.pt}
}

\maketitle
%Numbering
\acrodef{3GPP}{3rd Generation Partnership Project}
\acrodef{5G}{5th Generation Mobile Network}
\acrodef{6G}{6th Generation Mobile Network}
%-----A-----
\acrodef{AI}{Artificial Intelligence}
\acrodef{API}{Application Programming Interface}
\acrodef{AI4Net}{\ac{AI} for Networking}
\acrodef{AIDER}{Aerial Image Dataset for Emergency Response}
\acrodef{AMF}{Access and Mobility Management Function}
\acrodef{AIaaS}{Artificial Intelligence-as-a-Service}
\acrodef{AGORA}{Agentic Green Orchestration Architecture for Beyond 5G Networks}
\acrodef{IID}{Independent and Identically Distributed}
%-----B-----
\acrodef{B5G}{Beyond Fifth Generation}
\acrodef{BESS}{Berkeley Extensible Software Switch}
\acrodef{BPF}{Berkeley Packet Filter}
%-----C-----
\acrodef{CBR}{Constant Bit Rate}
\acrodef{CSV}{Comma-Separated Values}
\acrodef{CPU}{Central Processing Unit}
\acrodef{CNN}{Convolutional Neural Network}
\acrodef{CNNs}{Convolutional Neural Networks}
%-----D-----
\acrodef{DoS}{Denial of Service}
\acrodef{DDoS}{Distributed Denial of Service}
\acrodef{DDPG}{Deep Deterministic Policy Gradient}
\acrodef{DNN}{Deep Neural Network}
\acrodef{DPDK}{Data Plane Development Kit}
\acrodef{DRL}{Deep Reinforcement Learning}
\acrodef{DT}{Decision Tree}

%-----E-----
\acrodef{ETSI}{European Telecommunications Standards Institute}
\acrodef{eNWDAF}{Evolved Network Data Analytics Function}
\acrodef{EIF}{Energy Information Function}
\acrodef{E2E}{End-to-End}
%-----F-----
\acrodef{FIBRE}{Future Internet Brazilian Environment for Experimentation}
\acrodef{FPR}{False Positive Rate}
\acrodef{FAR}{Forwarding Action Rule}
%-----G-----
\acrodef{GNN}{Graph Neural Networks}
\acrodef{GPU}{Graphics Processing Unit}
\acrodef{GTP}{GPRS Tunnelling Protocol}
%-----H-----
\acrodef{HTM}{Hierarchical Temporal Memory}

%-----I-----
\acrodef{IAM}{Identity And Access Management}
\acrodef{ICMP}{Internet Control Message Protocol}
\acrodef{IID}{Informally, Identically Distributed}
\acrodef{IoE}{Internet of Everything}
\acrodef{IoT}{Internet of Things}
\acrodef{ITU}{International Telecommunication Union}
\acrodef{IBN}{Intent-Based Networking}
\acrodef{IoA}{Internet of Agents}
\acrodef{ISAC}{Integrated Sensing and Communication}
%-----J-----
%-----K-----
\acrodef{KNN}{K-Nearest Neighbors}
\acrodef{KPI}{Key Performance Indicator}
\acrodef{KPIs}{Key Performance Indicators}
%-----L-----
\acrodef{LSTM}{Long Short-Term Memory}
\acrodef{LLM}{Large Language Model}
%-----M-----
\acrodef{MAE}{Mean Absolute Error}
\acrodef{ML}{Machine Learning}
\acrodef{MLaaS}{Machine Learning as a Service}
\acrodef{MOS}{Mean Opinion Score}
\acrodef{MAPE}{Mean Absolute Percentage Error}
\acrodef{MSE}{Mean Squared Error}
\acrodef{MEC}{Multi-access Edge Computing}
\acrodef{mMTC}{Massive Machine Type Communications}
\acrodef{MFA}{Multi-factor Authentication}
\acrodef{MCP}{Model Context Protocol}
\acrodef{MBR}{Maximum Bit Rate}
%-----N-----
\acrodef{NWDAF}{Network Data Analytics Function}
\acrodef{Net4AI}{Networking for \ac{AI}}
\acrodef{NS}{Network Slicing}
\acrodef{NIC}{Network Interface Card}
%-----O-----
\acrodef{OSM}{Open Source MANO}
%-----P-----
\acrodef{PCA}{Principal Component Analysis}
\acrodef{PFCP}{Packet Forwarding Control Protocol}
\acrodef{PoC}{Proof of Concept}
\acrodef{PPV}{Positive Predictive Value}
\acrodef{PRB}{Physical Resource Block}
\acrodef{PDR}{Packet Detection Rule}
%-----Q-----
\acrodef{QoE}{Quality of experience}
\acrodef{QoS}{Quality of Service}
\acrodef{QER}{Quality of Service Enforcement Rule}
%-----R-----
\acrodef{RAM}{Random Access Memory}
\acrodef{RF}{Random Forest}
\acrodef{RL}{Reinforcement Learning}
\acrodef{RMSE}{Root Mean Square Error}
\acrodef{RNN}{Recurrent Neural Network}
\acrodef{RTT}{Round-Trip Time}
\acrodef{RAN}{Radio Access Network}
\acrodef{RIC}{RAN Intelligent Controller}

%-----S-----
\acrodef{SDN}{Software-Defined Networking}
\acrodef{SFI2}{Slicing Future Internet Infrastructures}
\acrodef{SLA}{Service-Level Agreement}
\acrodef{SON}{Self-Organizing Network}
\acrodef{SMF}{Session Management Function}
\acrodef{S-NSSAI}{Single Network Slice Selection Assistance Information}
\acrodef{SATSI}{Space-Air-Terrestrial-Sea Integrated Networks}

%-----T-----
\acrodef{TQFL}{Trust Deep Q-learning Federated Learning}
\acrodef{TEID}{Tunnel Endpoint Identifier}
\acrodef{TPS}{Transactions Per Second}
\acrodef{TPR}{True Positive Rate}
%-----U-----
\acrodef{UE}{User Equipment}
\acrodef{UEs}{User Equipments}
\acrodef{UPF}{User Plane Function}
\acrodef{UPFs}{User Plane Functions}
\acrodef{PDU}{Packet Data Unit}
\acrodef{URLLC}{Ultra-Reliable and Low Latency Communications}
\acrodef{UAV}{Unmanned Aerial Vehicle}
\acrodef{UAVs}{Unmanned Aerial Vehicles}
\acrodef{UDP}{User Datagram Protocol}
%-----V-----
\acrodef{VoD}{Video on Demand}
\acrodef{VR}{Virtual Reality}
\acrodef{VNF}{Virtual Network Function}
\acrodef{V2X}{Vehicle-to-Everything}

%-----W-----
%-----X-----
\acrodef{XR}{Extended Reality}
%-----Y-----
%-----Z-----
\acrodef{ZTN}{Zero-Touch Network}

\begingroup
\renewcommand\thefootnote{}
\footnotetext{\scriptsize
© 2026 IEEE. Personal use of this material is permitted. Permission from IEEE must be obtained for all other uses, in any current or future media, including reprinting/republishing this material for advertising or promotional purposes, creating new collective works, for resale or redistribution to servers or lists, or reuse of any copyrighted component of this work in other works.

Accepted author manuscript for publication in the Proceedings of the 2027 IEEE Consumer Communications \& Networking Conference (CCNC 2027), Las Vegas, Nevada, USA, January 9--12, 2027.
}
\addtocounter{footnote}{-1}
\endgroup

% Additional keys are supplied only if absent from the original acronym file.
\makeatletter
\newcommand{\providepaperacronym}[2]{\@ifundefined{fn@#1}{\newacro{#1}{#2}}{}}
\providepaperacronym{UE}{User Equipment}
\providepaperacronym{API}{Application Programming Interface}
\providepaperacronym{JSON}{JavaScript Object Notation}
\providepaperacronym{JSONL}{JSON Lines}
\providepaperacronym{DPU}{Data Processing Unit}
\providepaperacronym{eMBB}{enhanced Mobile Broadband}
\providepaperacronym{GTP-U}{GPRS Tunnelling Protocol for the User Plane}
\makeatother
\newacroplural{UE}[UEs]{User Equipment devices}
\begin{abstract}
The Third Generation Partnership Project (3GPP) Technical Specification 23.501 (Release 17) defines Maximum Bit Rate (MBR) enforcement at the User Plane Function (UPF) in fifth-generation (5G) networks. Supporting differentiated service-level agreements across the radio access and core networks requires the UPF to enforce MBR limits according to configured slice priorities. During overload, the challenge is to adapt these limits while preserving that differentiation and avoiding oscillations. Building on programmable enforcement and intelligent orchestration, we propose an agentic control loop combining UPF telemetry with locally hosted Large Language Model (LLM) selection from a finite set of admissible rate candidates. Deterministic logic validates each selection and derives Packet Forwarding Control Protocol (PFCP) updates. \textcolor{black}{The evaluated runs show no oscillations or fallbacks and unchanged highest-priority rate limits. These observations establish configuration preservation; delivered per-class QoS protection remains unmeasured.}
\end{abstract}


\section{Introduction}\label{sec:introduction}

Network slicing turns a shared mobile infrastructure into multiple service-specific networks, each bound to its own \ac{SLA} expectations~\cite{moreira2021,moreira2026-2}. This abstraction becomes increasingly consequential as mobile systems move toward \ac{6G}, where \ac{ISAC}, networked digital twins, AI-native automation, autonomous mobility, and immersive collaboration will depend on slice-level \ac{QoS} as an operational condition rather than a best-effort objective. At the same time, the commercial scale of \ac{5G} and the projected expansion of the network slicing market make differentiated enforcement \textcolor{black}{an operational requirement for operators}~\cite{Patil2024,Kochetkova2025}.

According to \ac{3GPP} TS~23.501, the \ac{UPF} enforces \ac{MBR} limits through \acp{QER}~\cite{3gpp23501}. Our control problem is to adjust these limits during overload, reducing lower-priority traffic while preserving the configured limits of the highest-priority slices.

\textcolor{black}{Existing approaches include predefined policies and learned controllers~\cite{Kyung2024,Damigos2025,Noe2026}. Our approach separates candidate selection from deterministic enforcement. A local \ac{LLM} selects among \ac{MBR} candidates constructed from live \ac{UPF} telemetry. We call these candidates \emph{\ac{SLA}-monotone} because, during overload, they can only maintain or reduce the affected class \ac{MBR} within the current phase. The controller validates the selection and derives the \ac{PFCP} action~\cite{moreira2026}.}

This paper proposes an agentic control loop that connects \ac{BESS} \ac{UPF} telemetry, \ac{SLA}-consistent candidate generation, \ac{LLM}-based candidate \textcolor{black}{selection, deterministic validation}, \ac{PFCP} actuation, and structured audit logging. \textcolor{black}{The evaluation establishes this integration under known traffic phases.}

Our contributions are as follows. (1) We design a per-window agentic QoS control loop for \ac{5G} network slicing over a real \ac{BESS} \ac{UPF} and \ac{DPDK} pipeline. (2) We introduce an SLA-monotone candidate policy and a derived action layer that \textcolor{black}{exclude within-phase rate increases during overload and derive executable actions independently of model-declared actions}. (3) We implement \ac{PFCP}-based per-session app-level \ac{MBR} modification for runtime enforcement. (4) We evaluate the controller on a two-site testbed \textcolor{black}{using static, threshold, greedy, and less constrained agentic controllers as reference operating points} under a four-phase overload schedule with ten independent runs per controller.

The remainder of this paper is organized as follows. Section~\ref{sec:related_work} reviews related work on \ac{QoS} enforcement and intelligent control for \ac{5G} network slicing. Section~\ref{sec:proposed_method} presents the proposed agentic control loop. Section~\ref{sec:experimental_testbed} describes our experimental testbed. Section~\ref{sec:results_and_discussion} reports the results and Section~\ref{sec:concluding_remarks} concludes the paper and outlines future work.

\section{Related Work}\label{sec:related_work}

\textbf{RAN slicing control}. Recent work on slice-aware \ac{QoS} control has first focused on radio resource allocation and \ac{RAN} slicing. Zipper addresses application-level throughput and latency assurance in \ac{5G} \ac{RAN} slicing through model predictive control and predictive bandwidth allocation~\cite{paper1}. AQPS targets multiservice \ac{RAN} resource allocation for eMBB, \ac{URLLC}, and best-effort slices through urgency-based priority scheduling~\cite{paper2}. DQN-ASRA jointly addresses slice prediction and radio block allocation for latency-sensitive services through enhanced Deep Q Network learning~\cite{paper5}. The O-RAN AIMLFW approach optimizes slice \ac{QoS} by predicting slice-level \ac{PRB} utilization and adjusting resource partitions through an rApp in the Non-Real-Time \ac{RIC}~\cite{paper4}. These works improve adaptive slice resource management, but they mainly operate in the \ac{RAN} domain rather than enforcing runtime \ac{MBR} decisions inside the \ac{UPF} data plane.

\textbf{User plane enforcement}.
A second line of work focuses on end-to-end service assurance and programmable user-plane enforcement. PCLANSA addresses proactive service assurance for end-to-end network slices through predictive \ac{VNF} scaling and closed-loop resource control~\cite{paper3}. P4-QUPF provides flow-level \ac{QoS} support in the \ac{UPF} through P4-based metering, dropping, and scheduling integrated with free5GC and \ac{PFCP} signaling~\cite{paper6}. 6GUPF improves scalable \ac{UPF} packet processing through DPU-based flow offload and hardware-assisted GTP-U handling~\cite{paper7}. The PSBA-XG user plane extension improves \ac{QoS} control by collecting \ac{UPF} usage reports and enabling \ac{PFCP}-based session reconfiguration and traffic rerouting~\cite{paper10}. These approaches move closer to the data plane. \textcolor{black}{Here, we focus on using \ac{UPF} telemetry and explicit per-session rate constraints before \ac{PFCP} actuation.}

\textbf{Agentic network automation}. A third group explores agentic and intent-driven intelligence for network management. The Agentic Intelligence assisted \ac{ML} fusion framework validates 6G slicing decisions against latency, throughput, and energy constraints through hybrid \ac{ML} agentic fusion~\cite{paper8}. The intent-based orchestration framework translates natural-language network intents into infrastructure and service actions through \ac{LLM}-based multi-agent orchestration and Infrastructure as Code~\cite{paper9}. These works demonstrate the potential of agentic control for network automation, but they do not address \ac{PFCP}-actuated \ac{UPF} \ac{QoS} enforcement under overload. In contrast, our work closes a telemetry-aware control loop over a real \ac{BESS} \ac{UPF} and \ac{DPDK} pipeline, where a locally hosted \ac{LLM} selects among \ac{SLA}-monotone \ac{MBR} candidates and a deterministic wrapper validates and applies the resulting action through \ac{PFCP}. \textcolor{black}{Action masking~\cite{revHuang2022Masking} and shielding~\cite{revAlshiekh2018Shielding} provide precedents for restricting learned decisions; our evaluation concerns their application to this implemented control loop.}

\begin{table}[htbp]
\centering
\caption{Comparison of State-of-the-Art Approaches.}
\label{tab:sota_survey}

\setlength{\tabcolsep}{1.3pt} 
\renewcommand{\arraystretch}{1.1} 

\scriptsize 

\resizebox{\columnwidth}{!}{%
\begin{tabular}{@{} l cccccccccc @{}}
\toprule
Approach & 
\makecell[c]{Slice\\QoS\\ctrl.} & 
\makecell[c]{Run-\\time\\adapt.} & 
\makecell[c]{Tele-\\metry\\loop} & 
\makecell[c]{Prio-\\rity\\protect.} & 
\makecell[c]{UPF\\data\\plane} & 
\makecell[c]{PFCP\\sess.\\mod.} & 
\makecell[c]{Prog.\\data\\plane} & 
\makecell[c]{LLM /\\agent\\ctrl.} & 
\makecell[c]{Struct.\\audit\\trail} & 
\makecell[c]{Real\\test-\\bed} \\ \midrule
\cite{paper1}  & \faCircle  & \faCircle  & \faCircle  & \faCircleO & \faCircleO & \faCircleO & \faCircleO & \faCircleO & \faCircleO & \faCircle  \\
\cite{paper2}  & \faCircle  & \faCircle  & \faCircleO & \faCircle  & \faCircleO & \faCircleO & \faCircleO & \faCircleO & \faCircleO & \faCircleO \\
\cite{paper3}  & \faCircle  & \faCircle  & \faCircle  & \faCircleO & \faCircleO & \faCircleO & \faCircleO & \faCircleO & \faCircleO & \faCircleO \\
\cite{paper4}  & \faCircle  & \faCircle  & \faCircle  & \faCircleO & \faCircleO & \faCircleO & \faCircleO & \faCircleO & \faCircleO & \faCircle  \\
\cite{paper5}  & \faCircle  & \faCircle  & \faCircleO & \faCircle  & \faCircleO & \faCircleO & \faCircleO & \faCircleO & \faCircleO & \faCircleO \\
\cite{paper6}  & \faCircle  & \faCircleO & \faCircleO & \faCircle  & \faCircle  & \faCircle  & \faCircleO & \faCircleO & \faCircleO & \faCircle  \\
\cite{paper7}  & \faCircle  & \faCircleO & \faCircleO & \faCircle  & \faCircleO & \faCircle  & \faCircleO & \faCircleO & \faCircleO & \faCircle  \\
\cite{paper8}  & \faCircle  & \faCircle  & \faCircleO & \faCircleO & \faCircleO & \faCircleO & \faCircleO & \faCircle  & \faCircleO & \faCircleO \\
\cite{paper9}  & \faCircleO & \faCircleO & \faCircleO & \faCircleO & \faCircleO & \faCircleO & \faCircleO & \faCircle  & \faCircleO & \faCircle  \\
\cite{paper10} & \faCircle  & \faCircle  & \faCircle  & \faCircle  & \faCircle  & \faCircleO & \faCircleO & \faCircleO & \faCircleO & \faCircle  \\
Ours        & \faCircle  & \faCircle  & \faCircle  & \textcolor{black}{P} & \faCircle  & \faCircle  & \faCircle  & \faCircle  & \faCircle  & \faCircle  \\ \bottomrule
\end{tabular}}
\vspace{0.3ex}
\par\smallskip\textcolor{black}{{\scriptsize \faCircle{}: fully supported; \faCircleO{}: not supported or not reported. \textcolor{black}{P: configured priority preservation only.}}}
\vspace{1.5ex}
\end{table}

\textcolor{black}{Table~\ref{tab:sota_survey} compares control objectives, enforcement mechanisms, and operational evidence. For our method, priority preservation refers to configured Gold \ac{MBR} limits; delivered per-class throughput, loss, and latency are not evaluated.}

\section{Proposed Method}\label{sec:proposed_method}

\begin{figure}[htp]
\centering
\definecolor{slGold}{HTML}{C99700}
\definecolor{slSilver}{HTML}{8C8C8C}
\definecolor{slBronze}{HTML}{A93226}
\begin{tikzpicture}
\pgfplotsset{
  every axis/.append style={
    width=\columnwidth, height=2.8cm,
    xmin=0, xmax=10, ticks=none,
    axis x line*={bottom, -stealth},
    axis x line*=bottom, axis y line*=left,
    axis line style={black!60, thin},
    clip=false,
  }
}
\begin{groupplot}[group style={group size=1 by 3, vertical sep=4pt}]

% ---------- (a) Static ----------
\nextgroupplot[ymin=0, ymax=345,
  legend columns=3,
  legend style={draw=none, fill=none, font=\normalsize,
                at={(0.5,1.06)}, anchor=south, column sep=6pt}]
\addplot[draw=none, fill=black!8, forget plot]
  coordinates {(4,345) (7,345)} \closedcycle;
\addplot[slGold, line width=1pt] coordinates
  {(0,200)(4,200)(4.3,182)(4.6,198)(4.9,180)(5.2,197)(5.5,181)
   (5.8,196)(6.1,180)(6.4,197)(6.7,182)(7,195)(7.3,200)(10,200)};
\addlegendentry{Gold}
\addplot[slSilver, line width=1pt] coordinates
  {(0,150)(4,150)(4.3,134)(4.6,148)(4.9,132)(5.2,147)(5.5,133)
   (5.8,148)(6.1,132)(6.4,147)(6.7,133)(7,146)(7.3,150)(10,150)};
\addlegendentry{Silver}
\addplot[slBronze, line width=1pt] coordinates
  {(0,100)(4,100)(4.25,250)(4.6,120)(4.95,252)(5.3,238)(5.65,251)
   (6.0,237)(6.35,120)(6.7,238)(7,248)(7.35,100)(10,100)};
\addlegendentry{Bronze}
\node[font=\normalsize, anchor=west] at (axis cs:0.15,310) {(a) Static};
\node[font=\normalsize, text=black!55] at (axis cs:5.5,310) {Overload};
\node[font=\scriptsize, text=black!55] at (axis cs:8.65,255) {loss hits Gold};

% ---------- (b) Threshold ----------
\nextgroupplot[ymin=0, ymax=300, ylabel={Throughput},
  ylabel style={font=\normalsize}]
\addplot[draw=none, fill=black!8, forget plot]
  coordinates {(4,300) (7,300)} \closedcycle;
\addplot[slGold, line width=1pt] coordinates {(0,200)(10,200)};
\addplot[slSilver, line width=1pt] coordinates
  {(0,150)(4,150)(4,150)(4.55,150)(4.55,150)(5.0,150)(5.0,150)(5.70,150)(5.70,50)
   (5.70,50)(6.20,50)(6.20,150)(6.55,150)(6.55,150)(7,150)(8,150)(9,150)(10,150)};
\addplot[slBronze, line width=1pt] coordinates
  {(0,100)(4,100)(4,18)(4.55,18)(4.55,80)(5.0,80)(5.0,18)(5.55,18)
   (5.55,80)(6.0,80)(6.0,18)(6.55,18)(6.55,80)(7.0,80)(7.0,100)(10,100)};
\node[font=\normalsize, anchor=west] at (axis cs:0.15,262) {(b) Threshold};
\node[font=\scriptsize, text=black!55] at (axis cs:8.55,46) {Bronze stop-and-go};

% ---------- (c) Agentic (desired) ----------
\nextgroupplot[ymin=0, ymax=300, xlabel={Time},
  xlabel style={font=\normalsize}]
\addplot[draw=none, fill=black!8, forget plot]
  coordinates {(4,300) (7,300)} \closedcycle;
\addplot[slGold, line width=1pt] coordinates {(0,200)(10,200)};
\addplot[slSilver, line width=1pt, smooth] coordinates
  {(0,150)(4,150)(4.7,129)(5.2,110)(6.8,137)(7.5,148)(8,150)(10,150)};
\addplot[slBronze, line width=1pt, smooth] coordinates
  {(0,100)(4,100)(4.8,62)(5.3,55)(6.8,55)(7.6,92)(8.1,100)(10,100)};
\node[font=\normalsize, anchor=west] at (axis cs:0.15,262) {(c) \textcolor{black}{Desired}};
\node[font=\scriptsize, text=black!55] at (axis cs:8.7,255) {Gold intact};
\node[font=\scriptsize, text=black!55] at (axis cs:8.7,44) {Bronze tapers};

\end{groupplot}
\end{tikzpicture}
\caption{Problem illustration for \ac{QoS} enforcement at the \ac{UPF}. \textcolor{black}{Gold, Silver, and Bronze curves are schematic, with no measured per-class \ac{QoS} values.}}
\label{fig:problem}
\end{figure}
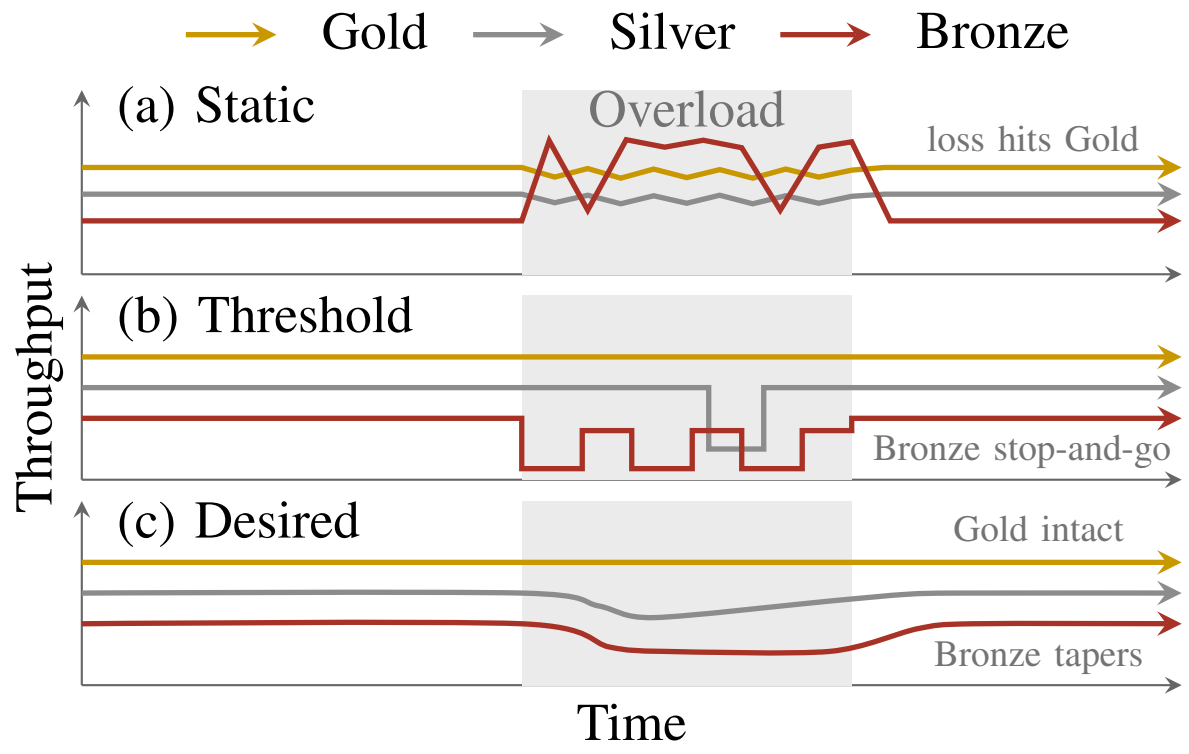

\textbf{Problem Setting}. \textcolor{black}{Figure~\ref{fig:problem} schematically motivates adapting lower-priority rates when traffic shares the same \ac{UPF}. Its desired service behavior is illustrative; the evaluation measures aggregate throughput and configured rate preservation.}

%\begin{figure}[htbp]
%  \centering
%  \includegraphics[width=0.55\columnwidth]{figures/method-problem-statement.pdf}
%  \caption{Problem statement for \ac{QoS} enforcement at the \ac{UPF}.}
%  \label{fig:problem}
%\end{figure}

%\textcolor{black}{Let $v_t$ denote the per-session \ac{MBR} vector and $v_0$ its initial configuration, representing the rate settings considered from the \ac{SLA}. Candidates $v\in\mathcal V_t$ satisfy $\sum_i v_i\le C$, with $C=800$~Mbps the configured rate budget. Gold stays at its initial limits, and $v_i\le(v_t)_i$ for the affected class defines \emph{\ac{SLA}-monotone} updates. For meter-red and pass packet counts $r_t,p_t$, the policing ratio is $\rho_t=r_t/(r_t+p_t)$, or zero when the denominator is zero. An oscillation is $v_t=v_{t-2}\ne v_{t-1}$ within one phase.}

%The closed loop in Figure~\ref{fig:method} runs the seven numbered steps once per window, summarized as Algorithm~\ref{alg:loop}. Steps 1 to 5 execute inside the controller; step 6 traverses the \ac{PFCP} control plane to the \ac{UPF}; step 7 returns to the controller.

\begin{figure}[htbp]
  \centering
  \includegraphics[width=0.6\columnwidth, angle=-90]{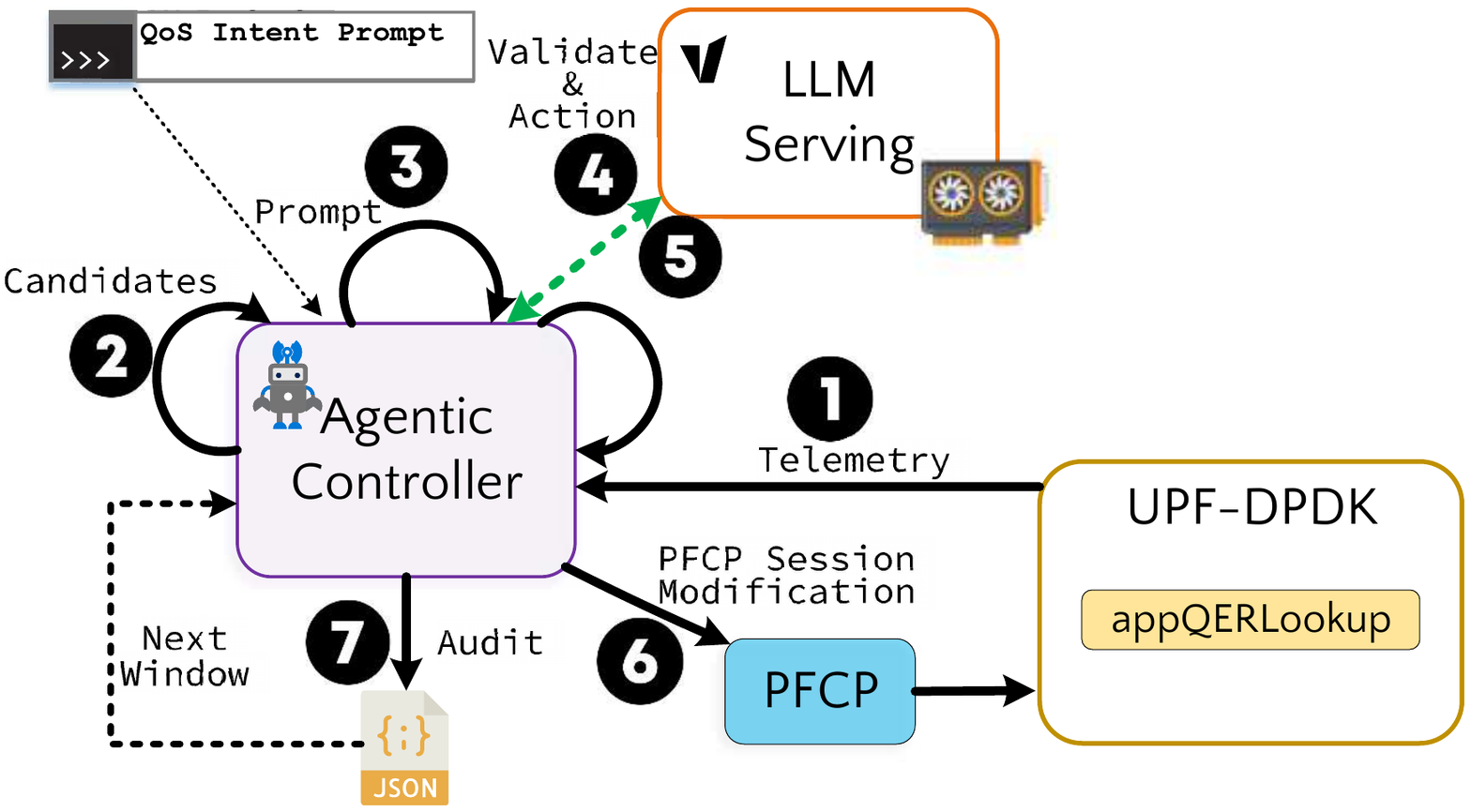}
\caption{Agentic control loop for \ac{QoS} enforcement in the \ac{UPF}.}
  \label{fig:method}
\end{figure}

To address this problem, Figure~\ref{fig:method} presents our control loop. Telemetry guides candidate construction, the \ac{LLM} selects a candidate, and the controller validates the selection before applying \ac{PFCP} updates. In Step 1, the controller pulls telemetry from sources, namely Prometheus and the \ac{BESS} interface on the modules along the pipeline. Then, the controller computes differences from the previous snapshot to produce per-window offered and delivered throughput, the app-\ac{QER} policing ratio, and the \ac{PDR}, \ac{FAR}, bad-route, and \ac{NIC} drop deltas.

In Step~2, the policy module uses the phase label supplied by the experimental schedule to identify the affected \textcolor{black}{service} class, while active overload requires a positive per-window app-\ac{QER} meter-red delta from Step~1. It then builds admissible \ac{MBR} vectors according to three rules. (i)~\emph{Priority pinning}: Gold slices remain fixed at their initial \ac{MBR} values. (ii)~\emph{Monotone tightening}: during overload, candidates may only maintain or reduce the affected class \ac{MBR} \textcolor{black}{within the current phase}. (iii)~\emph{Deterministic recovery}: during recovery, the candidate set collapses to the initial \ac{MBR} vector when needed. \textcolor{black}{The rate budget constrains configured \ac{MBR} sums.}

In Step~3, the controller builds the \ac{LLM} prompt using the current traffic phase, the telemetry deltas from Step~1, and the admissible candidates from Step~2. The prompt sets the \ac{QoS} priority order as Gold, Silver, then Bronze. It asks the model to select one candidate instead of generating \ac{MBR} values. \textcolor{black}{The required output includes a candidate identifier and audit fields. The controller retrieves the selected vector locally and compares it with the current vector to determine \textit{keep} or \textit{modify\_mbr}.}

The controller issues a single non-streaming chat completion to Qwen2.5-7B-Instruct served through vLLM in Step~4, using temperature 0 and a maximum of 300 output tokens. E4 and E4b use the same model and inference configuration, differing in the action space exposed to the model and in the deterministic derivation of executable actions. The response is parsed; only the candidate identifier is consumed by the runtime, while the textual reason is retained in the audit record for offline inspection\textcolor{black}{; it is a stored rationale}.

In Step 5, we check the chosen identifier against the candidate list, \textcolor{black}{configured rate budget}, and rule for the \ac{MBR} values. If any check fails, we keep the current vector and note the fallback. If all checks pass, we compare the chosen vector with the current vector. If they are the same, we retain it and stop the \ac{PFCP} path. If they are different, we change the \ac{MBR} and proceed to step 6.

In Step 6, a \textit{modify\_mbr} action is applied through a \ac{PFCP} Session Modification Request. The request updates the app-level \ac{MBR} of each session while keeping the session-level \ac{QER}, \ac{FAR}, \ac{PDR}, and \ac{TEID} states unchanged. Our pfcpsim sends the request to \ac{UPF}, and the new \ac{MBR} reaches the \ac{BESS} appQERLookup module, \textcolor{black}{before the next traffic window}.

Finally, in Step 7, the controller waits for the \ac{PFCP} response, then writes a single \textcolor{black}{\ac{JSONL}} record per window that includes the telemetry deltas, the candidate set, the raw and parsed \ac{LLM} response, the validation outcome, the derived action, the apply outcome, and the \ac{LLM} latency and token counts.

\section{Experimental Testbed}\label{sec:experimental_testbed}

\begin{figure}[htbp]
  \centering
  \includegraphics[width=0.7\columnwidth, angle=-90]{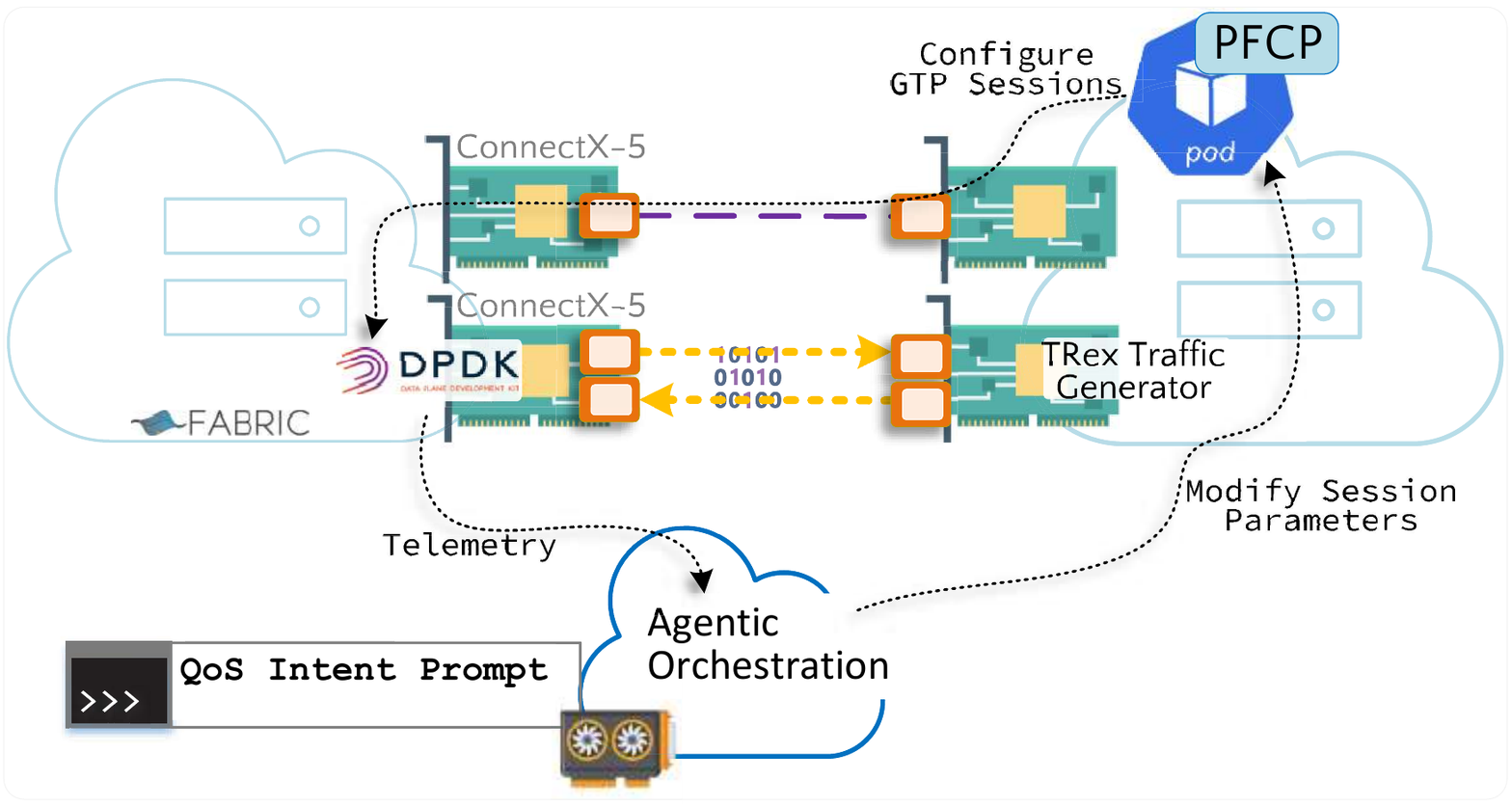}
  \caption{Our experimental testbed.}
  \label{fig:testbed}
\end{figure}

Our testbed\footnote{\scriptsize Available at https://github.com/romoreira/Agentic-B5G-QoS-Enforcement.} was deployed as a two-site FABRIC slice~\cite{Baldin2019}, as shown in Figure~\ref{fig:testbed}. One site hosts the \ac{UPF}, implemented with the OMEC BESS-UPF \ac{DPDK} pipeline and pfcpiface controller. The other site hosts the TRex traffic generator; we extended pfcpsim using Go routines to enable runtime \ac{MBR} modifications, alongside the Python runtime that executes the agentic control loop.

The data plane carries N3 GTP-U and N6 traffic over dedicated links, while \ac{PFCP} runs on a separate management interface. The \ac{LLM} endpoint runs outside the FABRIC slice and is accessed through a local API. Thus, each decision combines live \ac{UPF} telemetry, \ac{LLM} candidate selection, and \ac{PFCP} actuation over a real packet-processing pipeline.

The experiment uses five slices, two Gold, two Silver, and one Bronze. Table~\ref{tab:rate_settings} lists their initial \ac{MBR} values and the settings used by E2 and E3. Each TRex run follows four phases: nominal demand (A), Bronze overload (B), Silver overload (C), and recovery (D). E2 and E3 restore the initial settings in Phase~D.

We compare five controllers, E1 Static, E2 Threshold, E3 Greedy, E4 Agentic, and E4b Proposed. \textcolor{black}{E1--E3 span fixed-rate, threshold, and greedy policies; E4 assesses candidate restrictions and action derivation together.} Each controller is evaluated over ten independent runs using the same initial configuration, traffic schedule, and testbed conditions.

E1 keeps the initial settings throughout the run. E2 and E3 update once per phase. E4 and E4b can update after each 10~s traffic window. Telemetry collection, inference, and \ac{PFCP} actuation occur sequentially between windows. \textcolor{black}{The traffic-window duration is not the end-to-end control period. Different update schedules and admissible allocations prevent an isolated comparison of selector quality.}

\begin{table}[t]
\centering
\caption{\textcolor{black}{Configured \ac{MBR} per slice in Mbps. The initial configuration totals 800~Mbps across the five slices.}}
\label{tab:rate_settings}
\small
\setlength{\tabcolsep}{4pt}
\textcolor{black}{%
\begin{tabular}{llrrr}
\toprule
Controller & Phase & Gold & Silver & Bronze \\
\midrule
\multicolumn{2}{l}{Initial configuration}
                     & 200 & 150 & 100 \\
E2 Threshold & B      & 200 & 150 & 50  \\
             & C      & 200 & 120 & 50  \\
E3 Greedy    & B      & 180 & 140 & 160 \\
             & C      & 170 & 180 & 100 \\
\bottomrule
\end{tabular}%
}
\end{table}

\section{Results and Discussion}\label{sec:results_and_discussion}

\textcolor{black}{We compare E4b with the conventional policies E1, E2, and E3, and with the agentic variant E4. E4 permits rate relaxation during overload. E4b adds monotone candidate filtering and deterministic action derivation. Their comparison therefore evaluates both changes together.}

\begin{figure}[htbp]
  \centering
  \includegraphics[width=1\columnwidth]{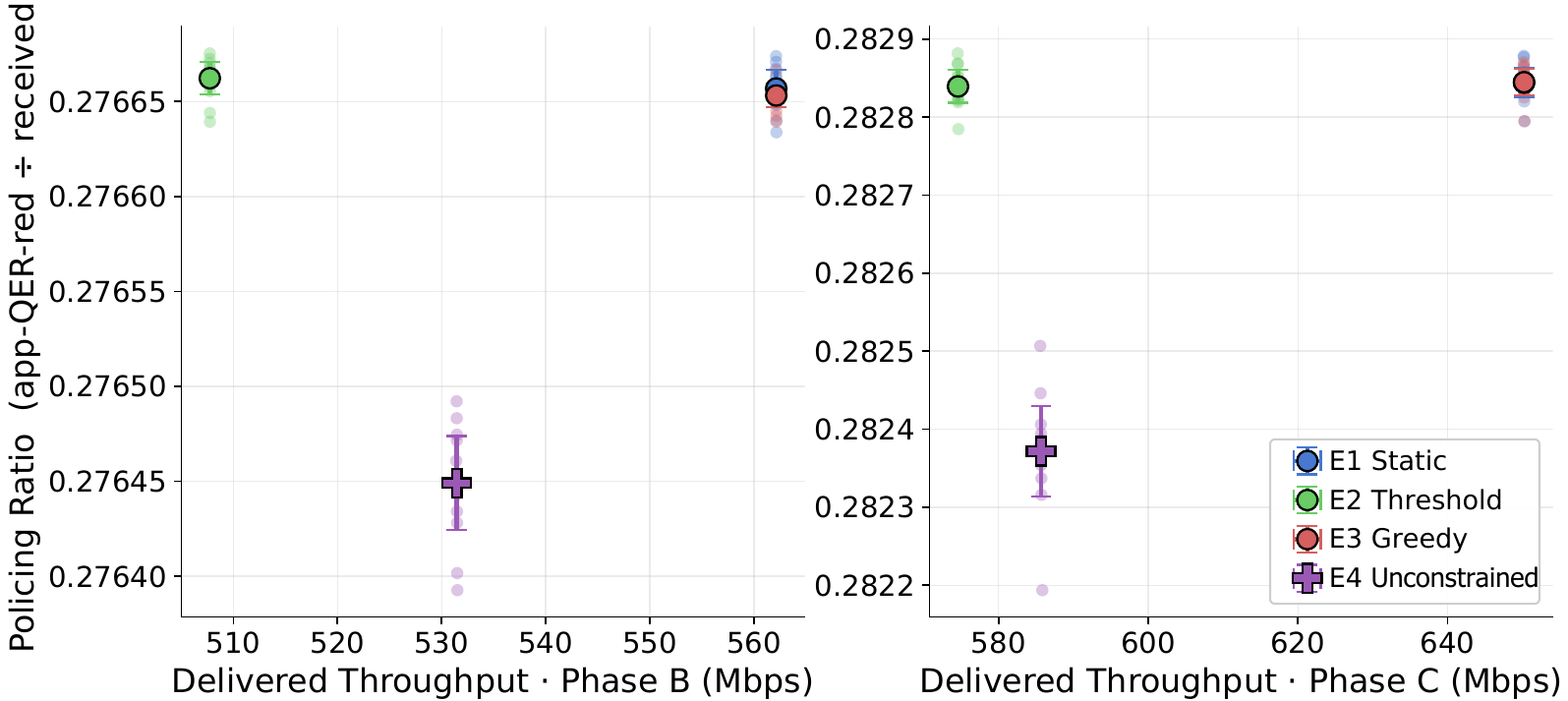}
  \caption{\textcolor{black}{Aggregate throughput and policing ratio during overload.}}
  \label{fig:pareto_throughput_policing}
\end{figure}

\textcolor{black}{Throughput is averaged within each phase and run. Error bars in Figure~\ref{fig:pareto_throughput_policing} show Student-$t$ 95\% confidence intervals across ten independent runs. The policing ratio is $\rho=r/(r+p)$, where $r$ and $p$ are the app-\ac{QER} meter-red and pass packet counts collected in each window. We set $\rho=0$ when both counts are zero.}

\textcolor{black}{Figure~\ref{fig:pareto_throughput_policing} and Table~\ref{tab:throughput_summary} summarize the aggregate results. In both overload phases, \textcolor{black}{both agentic controllers deliver} more throughput than Threshold and less than Static and Greedy.} \textcolor{black}{Without per-class measurements, the lower aggregate throughput cannot be attributed to a service-protection benefit.}

\begin{table}[t]
  \centering
  \caption{\textcolor{black}{Mean aggregate throughput during overload, in Mbps.}}
  \label{tab:throughput_summary}
  \small
  \setlength{\tabcolsep}{5pt}
  \begin{tabular}{lrr}
    \toprule
    Controller & Phase B & Phase C \\
    \midrule
    E1 Static    & 562.14 & 650.21 \\
    E2 Threshold & 507.73 & 574.58 \\
    E3 Greedy    & 562.14 & 650.22 \\
    E4 Agentic   & 531.47 & 585.69 \\
    E4b Proposed & 534.41 & 578.69 \\
    \bottomrule
  \end{tabular}
\end{table}

\textcolor{black}{A two-sided Mann--Whitney U test compares phase averages across independent runs.} The normal approximation with continuity correction gives $p=1.83\times10^{-4}$ for all six comparisons between E4 and the conventional policies. \textcolor{black}{With equal sample sizes, identical rank statistics yield identical $p$-values despite different throughput gaps. These tests do not assess E4b or robustness across traffic scenarios.}

\begin{figure}[htbp]
  \centering
  \includegraphics[width=\columnwidth]{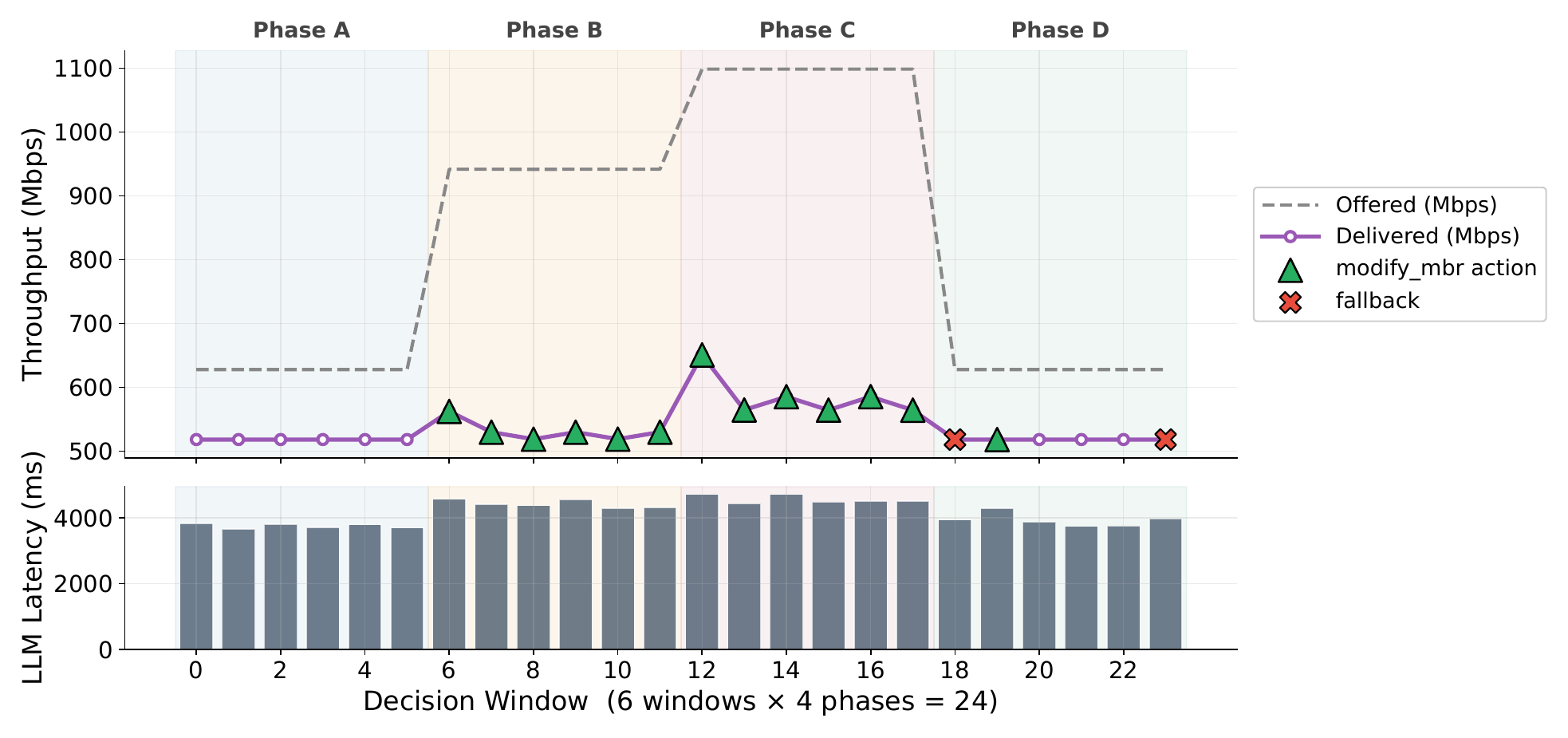}
  \caption{\textcolor{black}{Representative E4 behavior across the four traffic phases. The upper panel shows offered and delivered throughput, modifications, and fallbacks. The lower panel shows \ac{LLM} request latency.}}
  \label{fig:agentic_window}
\end{figure}

\textcolor{black}{The policies use different rate allocations. Static retains the initial limits, Threshold uses more restrictive settings, and Greedy reallocates capacity toward the overloaded class, including reductions in Gold limits. These allocation choices and the different update schedules must be considered when comparing throughput.} \textcolor{black}{Figure~\ref{fig:agentic_window} shows that E4 changes the \ac{MBR} allocation during operation. Most modifications occur during overload, accounting for 96.8\% of its updates. All Phase~A windows retain the current allocation. These records demonstrate runtime actuation.}

\begin{figure}[htbp]
  \centering
  \includegraphics[width=0.8\columnwidth]{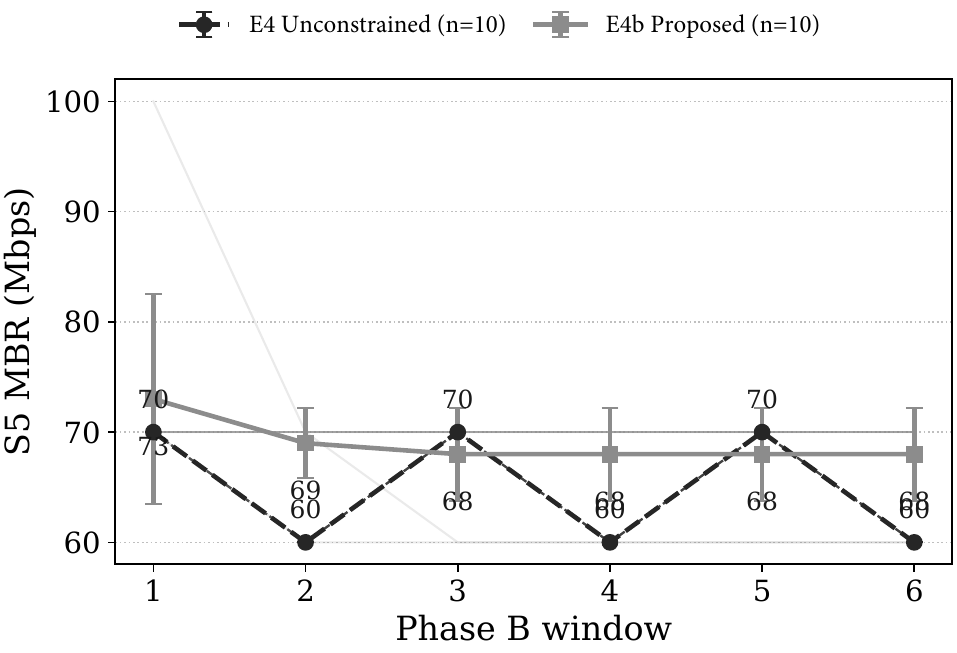}
  \caption{\textcolor{black}{Bronze slice \ac{MBR} in Phase~B under E4 and E4b.}}
  \label{fig:bronze_mbr_trajectory}
\end{figure}

\begin{figure}[htbp]
  \centering
  \includegraphics[width=0.75\columnwidth]{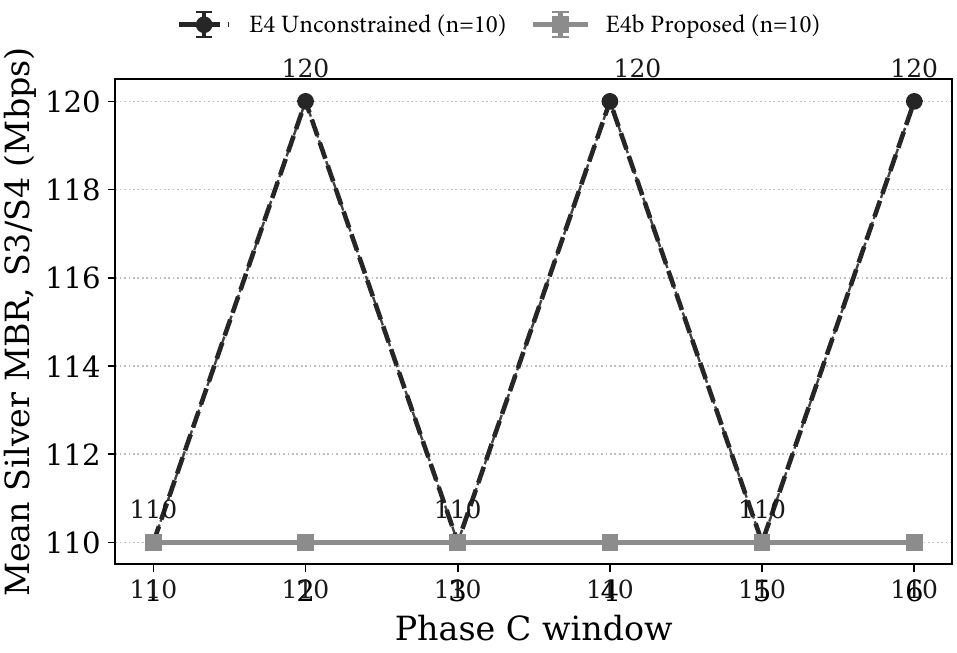}
  \caption{\textcolor{black}{Silver slice \ac{MBR} in Phase~C under E4 and E4b.}}
  \label{fig:silver_mbr_trajectory}
\end{figure}

\textcolor{black}{Figures~\ref{fig:bronze_mbr_trajectory} and~\ref{fig:silver_mbr_trajectory} show the limitation of E4. It alternates between reducing and increasing the Bronze allocation in Phase~B, with similar reversals for Silver in Phase~C. Each candidate can pass the local checks even though successive choices reverse earlier adjustments.} \textcolor{black}{E4b removes candidates that increase the affected class \ac{MBR} during overload within the same phase. \textcolor{black}{This within-phase monotonicity follows from deterministic filtering, independently of the selector.} The observed trajectories reach plateaus while Gold limits remain fixed.}

\begin{figure}[htbp]
  \centering
  \includegraphics[width=0.9\columnwidth]{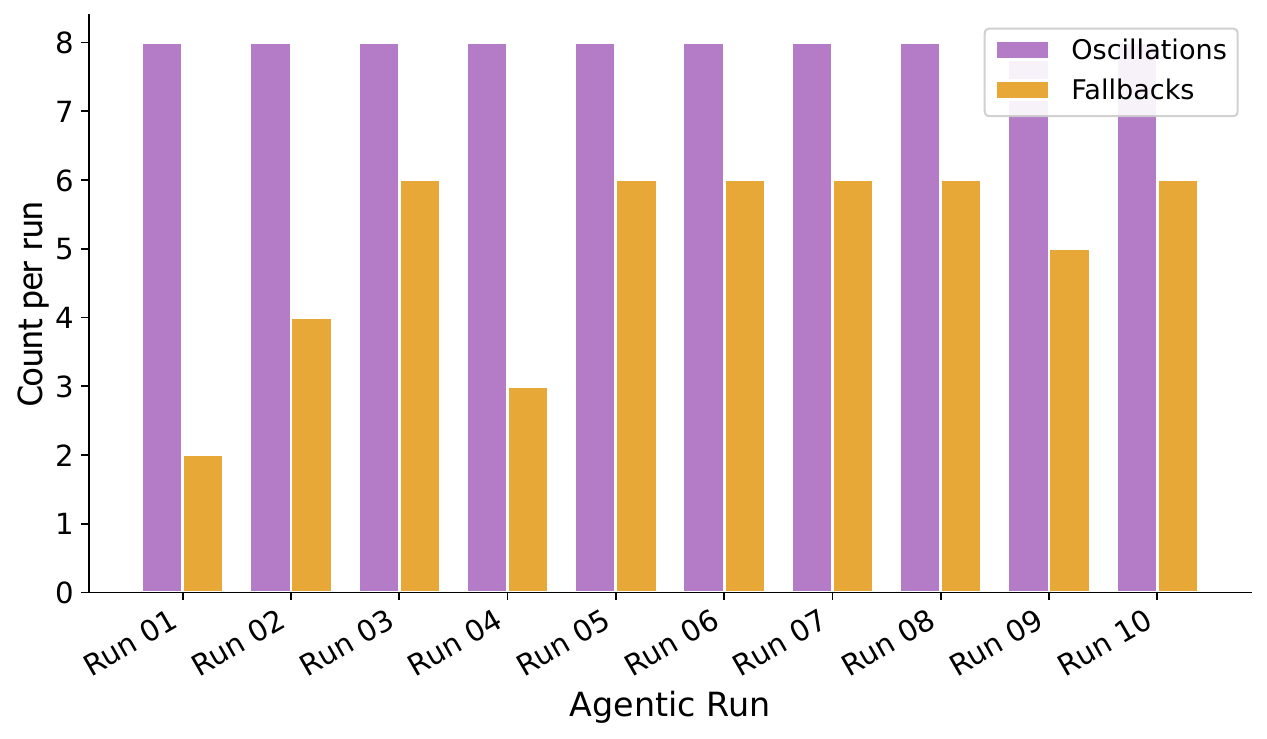}
  \caption{\textcolor{black}{Oscillations and fallbacks in the E4 runs.}}
  \label{fig:oscillations_fallbacks}
\end{figure}

\textcolor{black}{Figure~\ref{fig:oscillations_fallbacks} shows the events across E4 runs; Table~\ref{tab:control_outcomes} compares the totals with E4b. We count an oscillation when three consecutive \ac{MBR} vectors within one phase follow the pattern $x\rightarrow y\rightarrow x$, with $x\ne y$. A fallback retains the current vector after validation rejects the model output. E4 fallbacks occur mainly during recovery because the selected vector conflicts with the declared action. \textcolor{black}{E4b records neither event and applies fewer modifications in these runs; this observation provides no general reliability guarantee.} No \ac{PFCP} or datapath failures were observed.}

\begin{table}[t]
  \centering
  \caption{\textcolor{black}{Control events and maximum \ac{LLM} request latency across 240 windows per controller.}}
  \label{tab:control_outcomes}
  \small
  \setlength{\tabcolsep}{4pt}
  \textcolor{black}{%
  \begin{tabular}{lrr}
    \toprule
    Metric & E4 & E4b \\
    \midrule
    \ac{MBR} modifications & 124 & 32 \\
    Oscillations & 80 & 0 \\
    Fallbacks & 50 & 0 \\
    Maximum \ac{LLM} request (s) & 4.94 & 4.04 \\
    \bottomrule
  \end{tabular}%
  }
\end{table}

\begin{figure}[htbp]
  \centering
  \includegraphics[width=0.95\columnwidth]{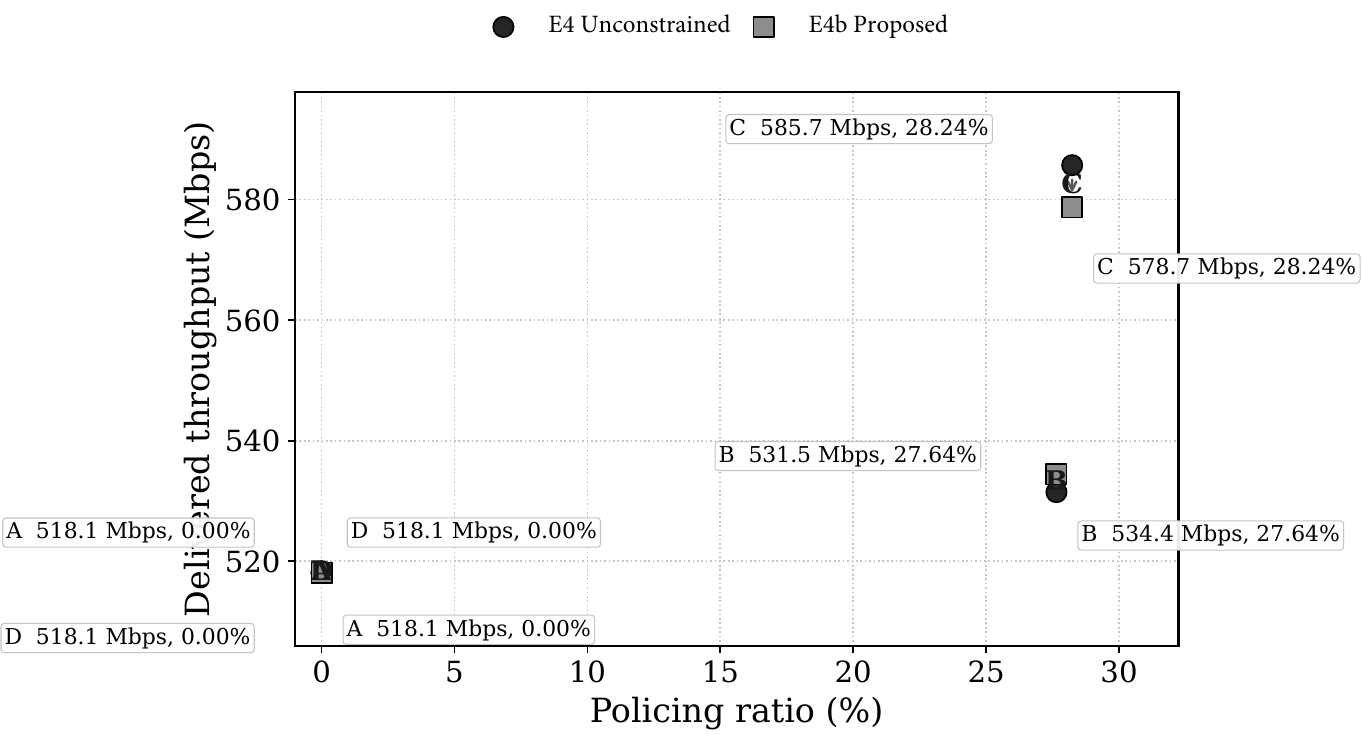}
  \caption{\textcolor{black}{Aggregate throughput and policing ratio under E4 and E4b across traffic phases.}}
  \label{fig:e4_e4b_efficiency}
\end{figure}

\textcolor{black}{Figure~\ref{fig:e4_e4b_efficiency} shows that E4 and E4b occupy nearby aggregate operating points. As Table~\ref{tab:throughput_summary} shows, E4b delivers slightly more throughput in Phase~B and slightly less in Phase~C. Its observed reduction in control events therefore comes with a modest throughput change in this scenario.}

\begin{figure}[htbp]
  \centering
  \includegraphics[width=0.90\columnwidth]{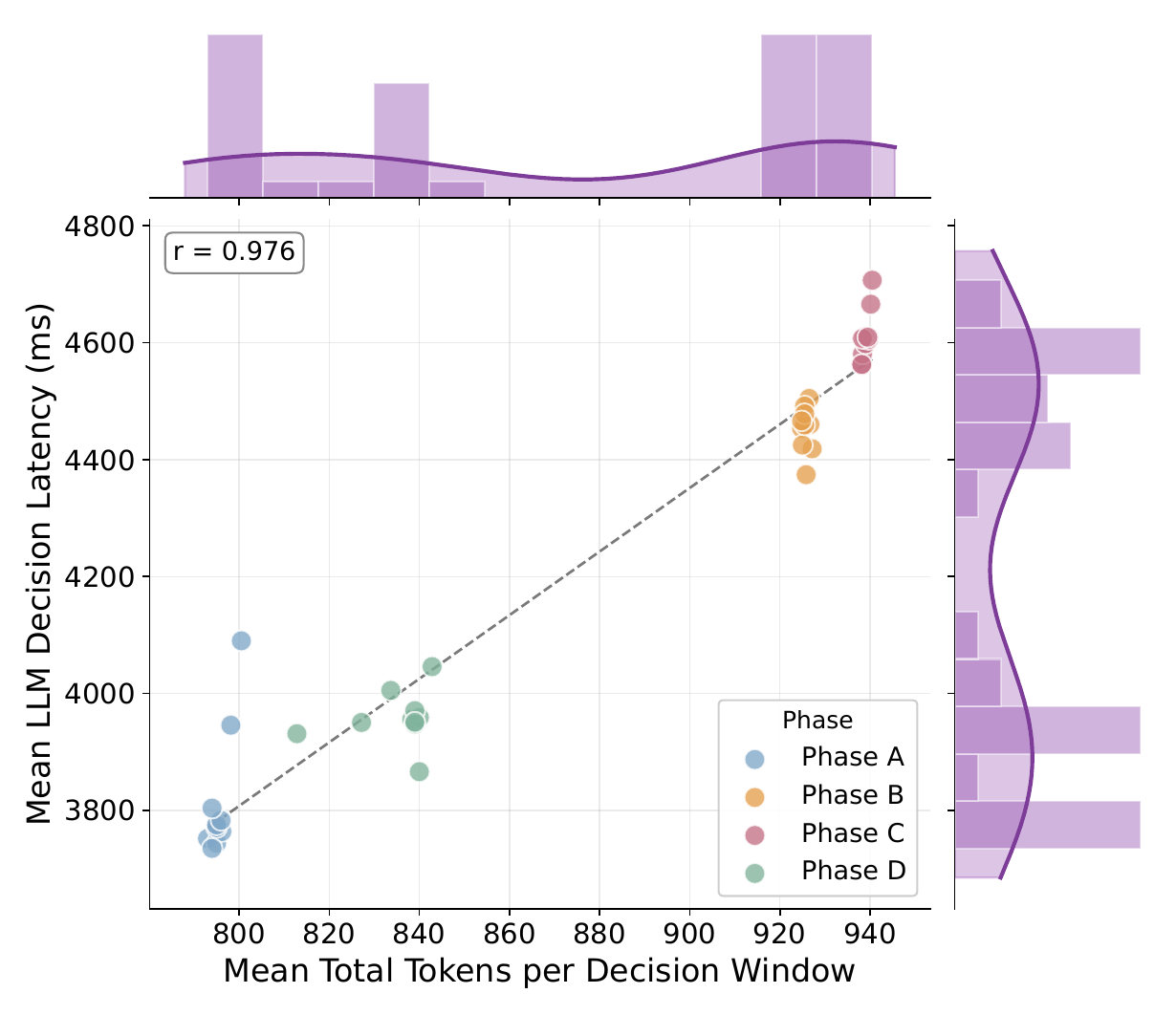}
  \caption{\textcolor{black}{Token volume and \ac{LLM} request latency.}}
  \label{fig:tokens_latency}
\end{figure}

\textcolor{black}{Figure~\ref{fig:tokens_latency} relates token volume to request latency. For E4b, the median is 2.72~s and the 95th percentile is 3.53~s. The maximum values appear in Table~\ref{tab:control_outcomes}. These measurements cover the model request and exclude telemetry collection and \ac{PFCP} actuation. \textcolor{black}{The median interval between E4b decision records is 44.4~s; this includes time outside the traffic window and is not an overload-to-enforcement measurement. Traffic continuity between windows and overload-to-enforcement delay are not reported.}}

\textcolor{black}{The results demonstrate a working integration of telemetry, model selection, deterministic validation, and \ac{PFCP} actuation. Gold limits remain unchanged across all 480 windows of E4 and E4b, confirming configuration preservation; delivered \ac{QoS} for each class was not measured.} \textcolor{black}{The incremental benefit of \ac{LLM} selection requires a deterministic selector with identical candidates, telemetry, constraints, and timing.}

\section{Concluding Remarks}\label{sec:concluding_remarks}

\textcolor{black}{This work implemented a telemetry-aware \ac{PFCP} control loop for adapting configured per-session \ac{MBR} limits on a real BESS-\ac{UPF} data plane. In the evaluated runs, deterministic candidate restrictions and action derivation preserved Gold limits and reduced control events. The evidence concerns configuration preservation and operational integration; delivered per-class \ac{QoS}, autonomous phase detection, and the incremental benefit of \ac{LLM} selection remain open. Future work should address these gaps through service-level measurements, unannounced traffic transitions, and a deterministic selector with identical candidates and timing.}

\section*{Acknowledgment}

The authors thank FAPEMIG (Grant \#APQ00923-24), FAPESP MCTIC/CGI Research project 2018/23097-3 - SFI2 - Slicing Future Internet Infrastructures, and Fundação para a Ciência e a Tecnologia (FCT) within the R\&D Unit Project Scope UID/00319/2025 - Centro ALGORITMI (ALGORITMI/UM) and the ANIMA Institute for supporting this work.

\bibliographystyle{IEEEtran}
% argument is your BibTeX string definitions and bibliography database(s)
\bibliography{references,agentic_revision_refs}

% Generated by IEEEtran.bst, version: 1.14 (2015/08/26)
\begin{thebibliography}{10}
\providecommand{\url}[1]{#1}
\csname url@samestyle\endcsname
\providecommand{\newblock}{\relax}
\providecommand{\bibinfo}[2]{#2}
\providecommand{\BIBentrySTDinterwordspacing}{\spaceskip=0pt\relax}
\providecommand{\BIBentryALTinterwordstretchfactor}{4}
\providecommand{\BIBentryALTinterwordspacing}{\spaceskip=\fontdimen2\font plus
\BIBentryALTinterwordstretchfactor\fontdimen3\font minus \fontdimen4\font\relax}
\providecommand{\BIBforeignlanguage}[2]{{%
\expandafter\ifx\csname l@#1\endcsname\relax
\typeout{** WARNING: IEEEtran.bst: No hyphenation pattern has been}%
\typeout{** loaded for the language `#1'. Using the pattern for}%
\typeout{** the default language instead.}%
\else
\language=\csname l@#1\endcsname
\fi
#2}}
\providecommand{\BIBdecl}{\relax}
\BIBdecl

\bibitem{moreira2021}
\BIBentryALTinterwordspacing
R.~Moreira, P.~F. Rosa, R.~L.~A. Aguiar, and F.~de~Oliveira~Silva, ``{NASOR: A network slicing approach for multiple Autonomous Systems},'' \emph{Computer Communications}, vol. 179, pp. 131--144, 2021. [Online]. Available: \url{https://www.sciencedirect.com/science/article/pii/S0140366421002917}
\BIBentrySTDinterwordspacing

\bibitem{moreira2026-2}
\BIBentryALTinterwordspacing
R.~Moreira, R.~Pasquini, J.~S. Martins, T.~C. Carvalho, and F.~{de Oliveira Silva}, ``{AI-driven orchestration at scale: Estimating service metrics on national-wide testbeds},'' \emph{Future Generation Computer Systems}, vol. 174, p. 107971, 2026. [Online]. Available: \url{https://www.sciencedirect.com/science/article/pii/S0167739X25002663}
\BIBentrySTDinterwordspacing

\bibitem{Patil2024}
A.~Patil, S.~Prabhu, and P.~Ashok, ``{5G Monetization Strategies and Business Models: An insightful Analysis for Telecommunications Service Providers},'' in \emph{2024 5th International Conference on Data Intelligence and Cognitive Informatics (ICDICI)}, 2024, pp. 16--23.

\bibitem{Kochetkova2025}
I.~Kochetkova, K.~Leonteva, I.~Ghebrial, and A.~Khakimov, ``{Dynamic Spectrum Management for 5G NR Network Slicing: Frequency Reallocation Ensuring Slice Isolation and Revenue Maximization},'' \emph{IEEE Access}, vol.~13, pp. 208\,414--208\,438, 2025.

\bibitem{3gpp23501}
{3GPP}, ``System architecture for the {5G} system ({5GS}),'' 3rd Generation Partnership Project, Technical Specification 23.501, Oct. 2024, version 17.14.0, Release 17.

\bibitem{Kyung2024}
Y.~Kyung, D.~Lee, J.~Lee, H.~Ko, S.~Oh, J.~Lee, and Y.~Kim, ``{PDU Set-Based Traffic Handling for XR Service Support in 5G User Plane},'' in \emph{2024 15th International Conference on Information and Communication Technology Convergence (ICTC)}, 2024, pp. 2155--2156.

\bibitem{Damigos2025}
G.~Damigos, E.~Pagliari, S.~Sandberg, and G.~Nikolakopoulos, ``{5G-enabled robots: Differentiated connectivity for varying mission requirements through dynamic QoS},'' \emph{IEEE Transactions on Vehicular Technology}, pp. 1--16, 2025.

\bibitem{Noe2026}
N.~M. Yungaicela-Naula, V.~Sharma, and S.~Scott-Hayward, ``{RSLAQ- A Robust SLA-driven 6G O-RAN QoS xApp Using Deep Reinforcement Learning},'' \emph{IEEE Transactions on Mobile Computing}, pp. 1--17, 2026.

\bibitem{moreira2026}
R.~Moreira, L.~F. Rodrigues~Moreira, M.~Leone Maciel~Peixoto, and F.~D. Oliveira~Silva, ``{AGORA: Agentic Green Orchestration Architecture for Beyond 5G Networks},'' \emph{IEEE Access}, vol.~14, pp. 43\,519--43\,532, 2026.

\bibitem{paper1}
\BIBentryALTinterwordspacing
A.~Balasingam, M.~Kotaru, and P.~Bahl, ``{Application-Level} service assurance with 5g {RAN} slicing,'' in \emph{21st USENIX Symposium on Networked Systems Design and Implementation (NSDI 24)}.\hskip 1em plus 0.5em minus 0.4em\relax Santa Clara, CA: USENIX Association, Apr. 2024, pp. 841--857. [Online]. Available: \url{https://www.usenix.org/conference/nsdi24/presentation/balasingam}
\BIBentrySTDinterwordspacing

\bibitem{paper2}
A.~Yaqoob and G.-M. Muntean, ``{An Adaptive QoS-Aware Priority Scheduling Solution for Dynamic Radio Resource Allocation in Multiservice 5G Network Slicing Environments},'' \emph{IEEE Internet of Things Journal}, vol.~13, no.~6, pp. 12\,244--12\,260, 2026.

\bibitem{paper5}
\BIBentryALTinterwordspacing
M.~Ya{\u{g}}c{\i}o{\u{g}}lu, ``{Enhanced Deep Reinforcement Learning-Driven Adaptive Network Slicing and Resource Allocation for URLLC in 5G Networks},'' \emph{Journal of Network and Systems Management}, vol.~34, no.~1, p.~33, Dec 2025. [Online]. Available: \url{https://doi.org/10.1007/s10922-025-10009-2}
\BIBentrySTDinterwordspacing

\bibitem{paper4}
M.~H. Rahman, A.~Jain, A.~Da~Silva, S.~Singh, M.~Roy~Chowdhury, H.~S. Dhillon, and M.~Hong, ``{Leveraging O-RAN SC AI/ML Framework \& Non-RT RIC for AI-Driven Network Slice QoS Optimization},'' in \emph{2024 IEEE Globecom Workshops (GC Wkshps)}, 2024, pp. 1--6.

\bibitem{paper3}
N.~Phuc~Tran, O.~Delgado, and B.~Jaumard, ``{Proactive Service Assurance in 5G and B5G Networks: A Closed-Loop Algorithm for End-to-End Network Slices},'' \emph{IEEE Transactions on Network and Service Management}, vol.~23, pp. 668--680, 2026.

\bibitem{paper6}
\BIBentryALTinterwordspacing
M.~S. Iqbal, M.~Elangovan, C.~Chen, and J.-C. Chen, ``{P4-QUPF: A P4-Based QoS Framework for 5G-UPF},'' \emph{International Journal of Network Management}, vol.~35, no.~6, p. e70028, 2025, e70028 NEM-24-1822.R1. [Online]. Available: \url{https://onlinelibrary.wiley.com/doi/abs/10.1002/nem.70028}
\BIBentrySTDinterwordspacing

\bibitem{paper7}
R.~A. Bakar, A.~S.~T. Ibrahim, F.~Paolucci, A.~Sgambelluri, P.~Castoldi, F.~Cugini, and J.~J. Vegas~Olmos, ``6gupf: A dpu-based programmable user plane function for enhanced flow-based qos,'' in \emph{GLOBECOM 2025 - 2025 IEEE Global Communications Conference}, 2025, pp. 793--798.

\bibitem{paper10}
V.~Romain, T.~Lejkin, S.~Rovedakis, and S.~Secci, ``{A Data-Centric Approach for User Plane Control to Improve QoS for Beyond 5G Systems},'' in \emph{2025 14th IFIP/IEEE International Conference on Performance Evaluation and Modeling in Wired and Wireless Networks (PEMWN)}, 2025, pp. 1--8.

\bibitem{paper8}
S.~Vashisht, S.~Rani, and H.~Feng, ``{Multimodal and Agentic Intelligence–Driven ML Fusion for Sustainable 6G Network Slicing},'' \emph{IEEE Open Journal of the Communications Society}, vol.~7, pp. 1643--1653, 2026.

\bibitem{paper9}
D.~Brodimas, A.~Birbas, D.~Kapolos, and S.~Denazis, ``{Intent-Based Infrastructure and Service Orchestration Using Agentic-AI},'' \emph{IEEE Open Journal of the Communications Society}, vol.~6, pp. 7150--7168, 2025.

\bibitem{revHuang2022Masking}
S.~Huang and S.~Onta{\~n}{\'o}n, ``A closer look at invalid action masking in policy gradient algorithms,'' \emph{The International FLAIRS Conference Proceedings}, vol.~35, 2022.

\bibitem{revAlshiekh2018Shielding}
M.~Alshiekh, R.~Bloem, R.~Ehlers, B.~K{\"o}nighofer, S.~Niekum, and U.~Topcu, ``Safe reinforcement learning via shielding,'' \emph{Proceedings of the AAAI Conference on Artificial Intelligence}, vol.~32, no.~1, 2018.

\bibitem{Baldin2019}
I.~Baldin, A.~Nikolich, J.~Griffioen, I.~I.~S. Monga, K.-C. Wang, T.~Lehman, and P.~Ruth, ``{FABRIC: A National-Scale Programmable Experimental Network Infrastructure},'' \emph{IEEE Internet Computing}, vol.~23, no.~6, pp. 38--47, 2019.

\end{thebibliography}

\end{document}